\documentstyle[12pt,psfig]{article}

\begin{document}

\newcommand{\psla}{p\hspace{-0.45em}/}
\def\be{\begin{eqnarray}}
\def\ee{\end{eqnarray}}
\def\ep{\varepsilon}

\thispagestyle{empty}
\begin{flushright}
{\bf TTP96-52}\\ 
{hep-ph/9611299}\\
{\bf November 1996}\\
\end{flushright}
\vspace*{20mm}
\begin{center}
  \begin{large}
On the Dipole Moments of Fermions at Two Loops\\
  \end{large}
  \vspace{0.5cm}
  \begin{large}
Andrzej Czarnecki and Bernd Krause
 \\
  \end{large}
  \vspace{5mm}
{\em Institut f\"ur Theoretische Teilchenphysik, \\
Universit\"at Karlsruhe,
D-76128 Karlsruhe, Germany} \\[4mm]

\vspace{40mm}
  {\bf Abstract}\\
\vspace{0.3cm}
\noindent
\begin{minipage}{12.5cm}
Complete two-loop electroweak corrections to leptonic
anomalous magnetic moments have been calculated recently using
asymptotic expansions of Feynman diagrams.  Techniques developed in
that context can also be applied to the electric dipole moments (EDM) 
of fermions.   In the Standard Model the EDM was proved to vanish at two
loops.  In this talk we discuss a simplification of that proof.
\end{minipage}
\end{center}
\hfill

\newpage
\setcounter{page}{1}

%%%%%%%%%%%%%%%%%%%%%%%%%%%%%%%%%%%%%%%%%%%%%%%%%%%%%%%%%%%%
\noindent

\section{Introduction}
In a recent calculation of two-loop
electroweak corrections to $g-2$ of leptons \cite{CKM95} we have 
applied computing techniques based on asymptotic expansion of the
Feynman diagrams \cite{Smi94}.  Similar tools have many other important
applications, especially in low energy physics.  In the present talk we
focus on the two-loop contributions to the electric dipole moments
(EDM) of fermions.

Precise measurements of 
the electric dipole moment of the neutron and the electron
impose severe constraints on extensions of the 
Standard Model (SM) and even on the SM itself ($\theta$-term in
the QCD Lagrangian) \cite{Barr89}. 
If a non-vanishing EDM of an elementary particle were found, it would be the
first manifestation of CP-violation outside the $K^0-{\bar K^0}$ system
and could provide essential information about the nature of
CP-violation. 
For this reason two decades ago several groups started the
theoretical investigation of EDMs in the SM \cite{EDM76}. They found that EDMs
vanish trivially at the one-loop level in the SM because
only the absolute values of the Cabibbo-Kobayashi-Maskawa
(CKM) matrix elements enter the relevant
amplitudes, not permitting for a CP-violating complex phase. However,
at the two-loop level (with two virtual $W$ bosons) they found a
quark flavor structure rich enough to generate a non-vanishing EDM of the
neutron.

In 1978 a more detailed analysis of Shabalin \cite{S78} and Donoghue
\cite{D78} 
showed  that the EDM of  both the neutron and the
electron must vanish identically at the two-loop level.
This result is somewhat surprising since it could not be traced back
to any symmetry which would enforce zero EDM. 
Subsequent studies \cite{EG79,S80,Kr86} found that 
QCD corrections may generate a
non-vanishing EDM of the order of $G_F^2 \alpha_s$.
These papers, however, differ strongly in the predictions of the
size of the three-loop contributions.

The proofs of the vanishing of EDM are based on cancellations among
certain groups of diagrams.  We noticed that these cancellations are
stronger than previously believed, that is they occur among smaller
subsets of diagrams.  Before presenting our results we summarize in
the next section the original argument of Shabalin's proof.

We believe the simplifications we found contribute to a better
understanding of the perturbative contributions to EDM and may
eventually help in the clarification of the three-loop value of the
quark EDM. 

\section{Shabalin's Argument \protect\cite{S80}}

In the unitary gauge the diagrams contributing to EDM at the two-loop
level are obtained by attaching an external photon to every internal
line of the diagram depicted in fig. 1.\\[4mm]

%%%%%%%%%%%%%%%%%%%%%%%%%%%%%%%%%%%%%%%%%%%%%%%%%%%%%%%%%%%%
\vspace{0.1cm}
\begin{minipage}{16.cm}
\hspace*{.8cm}
\[
\mbox{
\hspace*{-30mm}
\begin{tabular}{ccc}
\psfig{figure=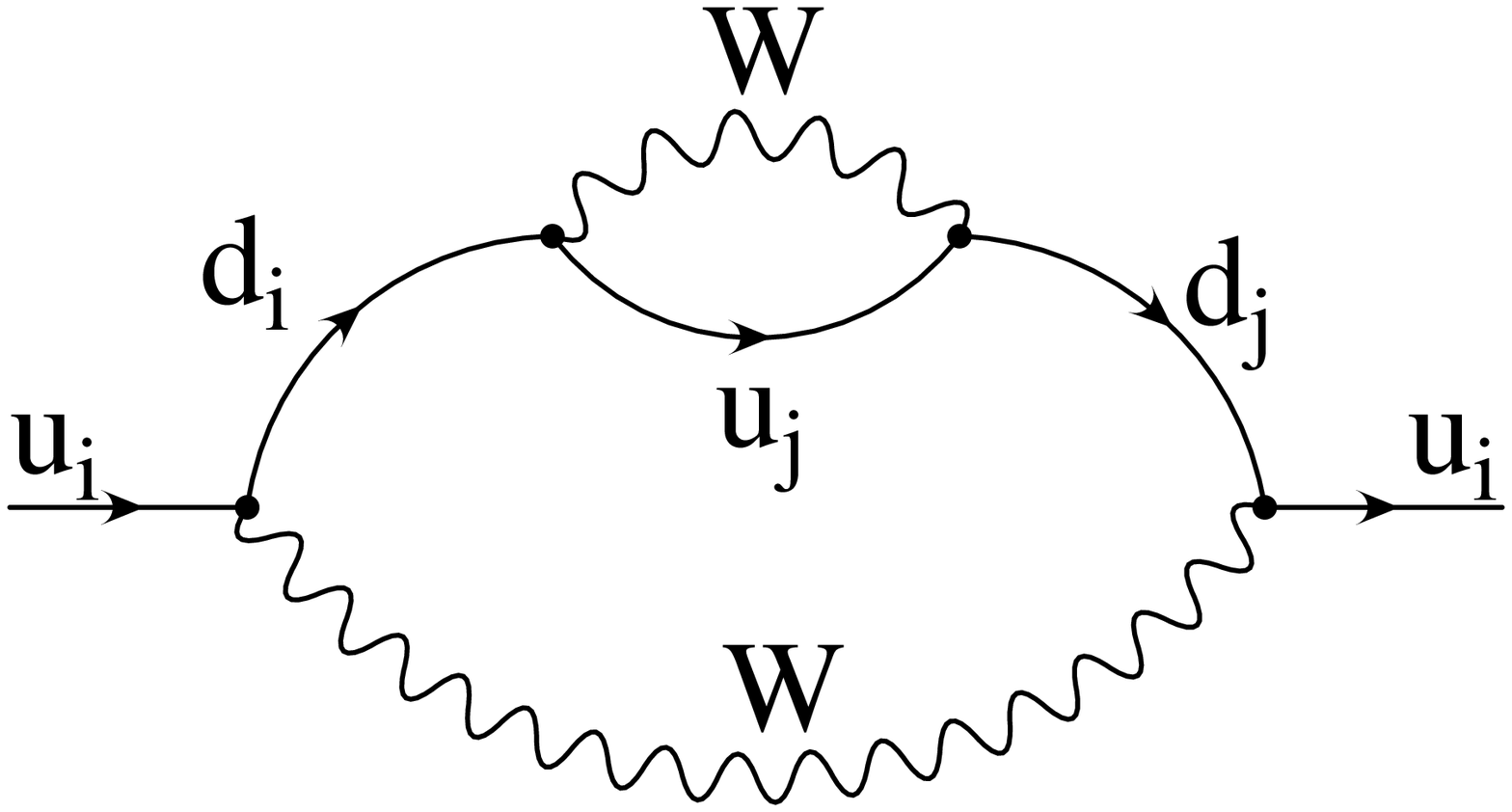,width=25mm,bbllx=210pt,%
bblly=410pt,bburx=630pt,bbury=550pt} 
&\hspace*{2cm}
\psfig{figure=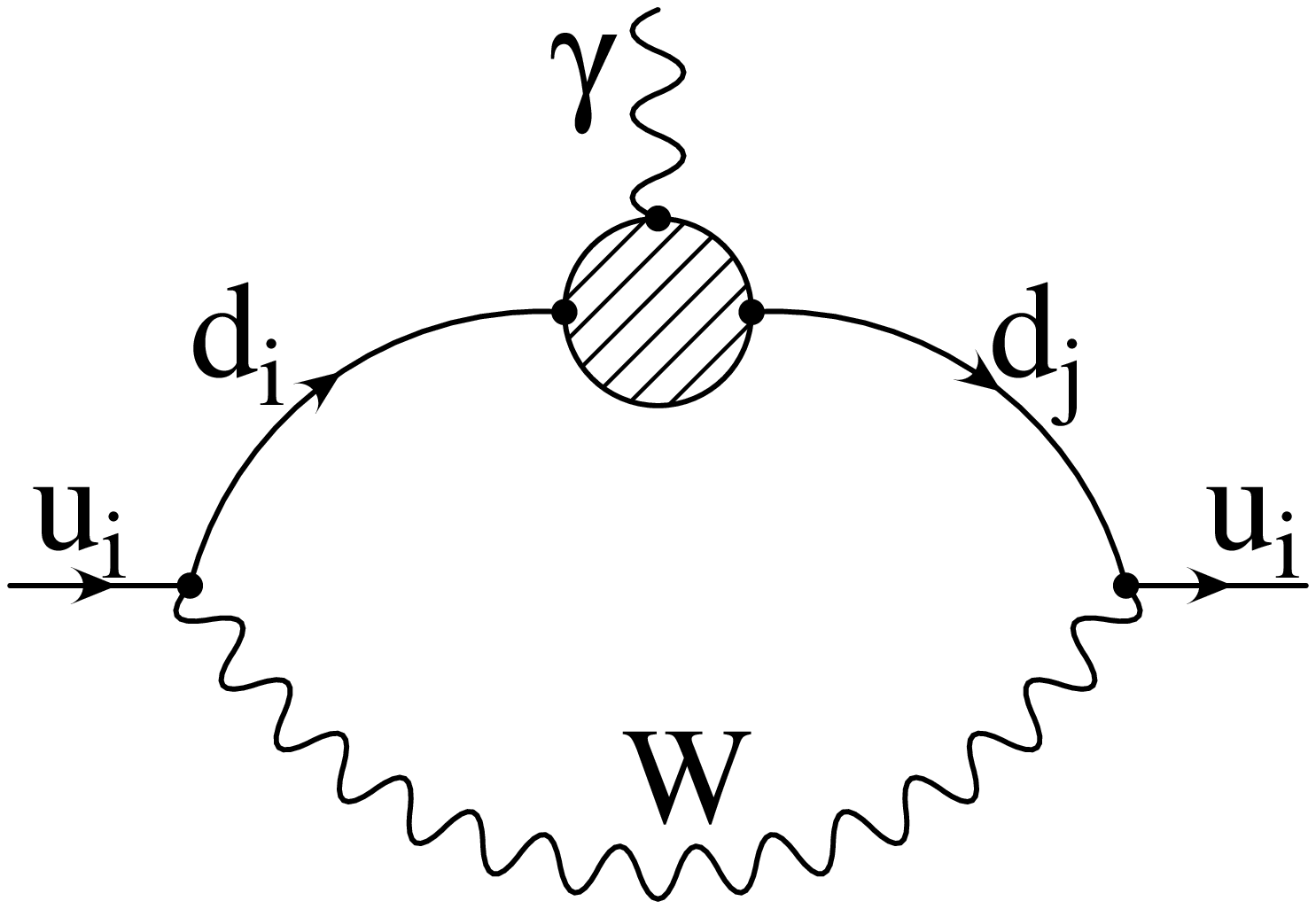,width=25mm,bbllx=210pt,%
bblly=410pt,bburx=630pt,bbury=550pt}
&\hspace*{.8cm}
\psfig{figure=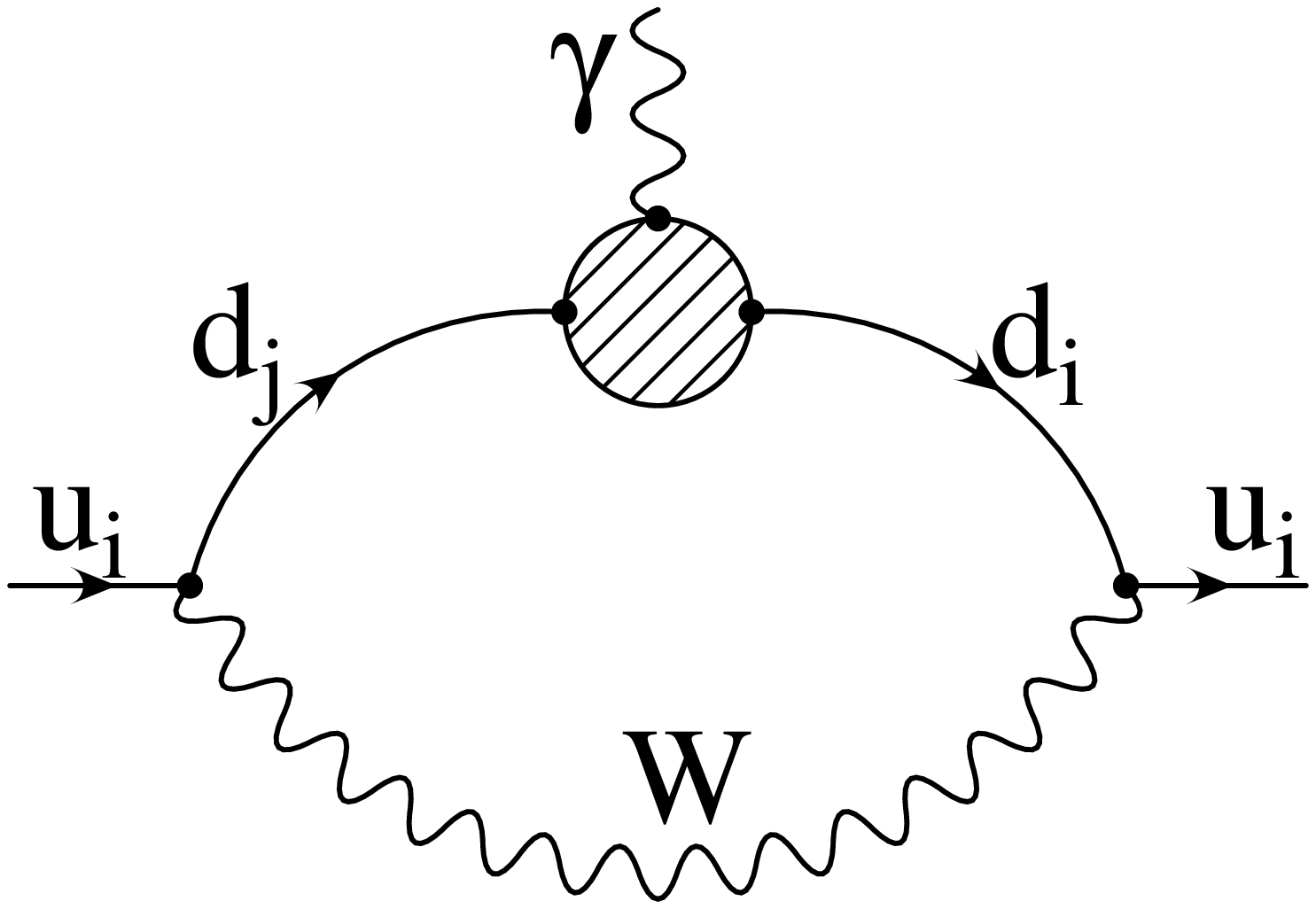,width=25mm,bbllx=210pt,%
bblly=410pt,bburx=630pt,bbury=550pt}
\\[9mm]
\end{tabular}}
\]
\end{minipage}

\noindent 
\hspace{14mm}Fig. (1)\hspace{34mm} Fig. (2a)\hspace{20mm} Fig. (2b)\\
\vspace{.1cm}
%%%%%%%%%%%%%%%%%%%%%%%%%%%%%%%%%%%%%%%%%%%%%%%%%%%%%%%%%%%%

These diagrams possess a complex CKM phase, necessary to generate EDM
(a T violating observable) if all quarks $u_i$, $d_i$ are different.

We now divide the diagrams generated from fig.~1 into two sets:
the first set consists of only one diagram, where the photon couples to the
outer $W$ line; one can check by an explicit calculation that the part
of this diagram which contains $\gamma_5$ vanishes, and therefore it
does not contribute to EDM.  The sum of vertex and selfenergy
counterterms for diagrams obtained from fig.~1 gives no contribution
to EDM either.

In the rest of this paper we will focus on the remaining set of
diagrams. It can be viewed as a one-loop diagram with
an insertion of a flavor-changing effective vertex $F_1 F_2\gamma$ 
with off-shell fermions $F_{1,2}$ (see fig.~2a). This effective vertex is
given by a sum  of one-loop diagrams in fig. 3.

%%%%%%%%%%%%%%%%%%%%%%%%%%%%%%%%%%%%%%%%%%%%%%%%%%%%%%%%%%%%
\vspace{0.1cm}
\begin{minipage}{12.cm}
\hspace*{-.8cm}
\[
\mbox{
\hspace*{-10mm}
\begin{tabular}{cccccccc}
$\Gamma^{F_1 F_2\gamma}=$
&\hspace*{3mm}
\psfig{figure=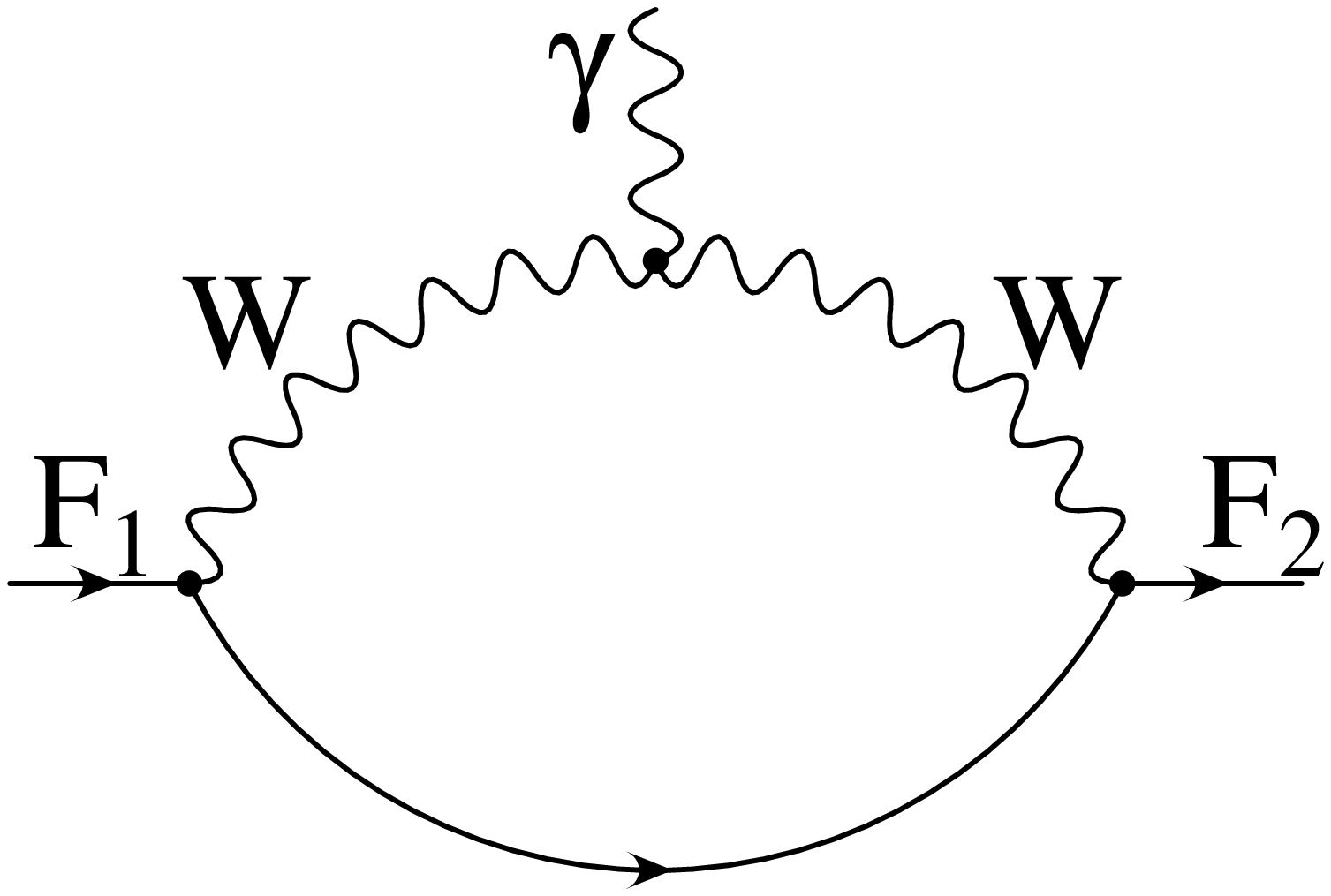,width=21mm,bbllx=210pt,%
bblly=410pt,bburx=630pt,bbury=550pt} 
&\hspace*{-6mm}$+$\hspace*{4mm}&
\psfig{figure=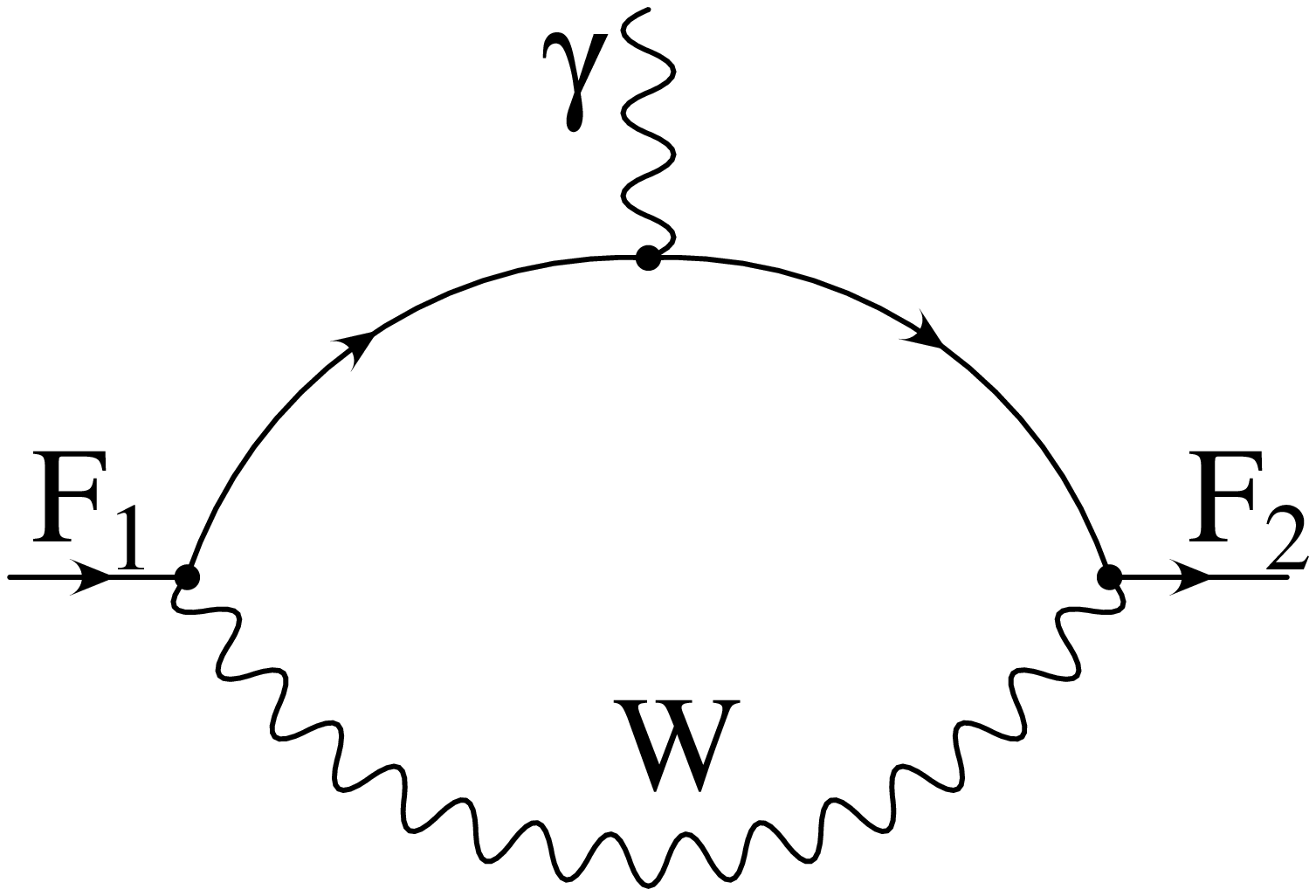,width=21mm,bbllx=210pt,%
bblly=410pt,bburx=630pt,bbury=550pt}
&\hspace*{-6mm}$+$\hspace*{4mm}&
\psfig{figure=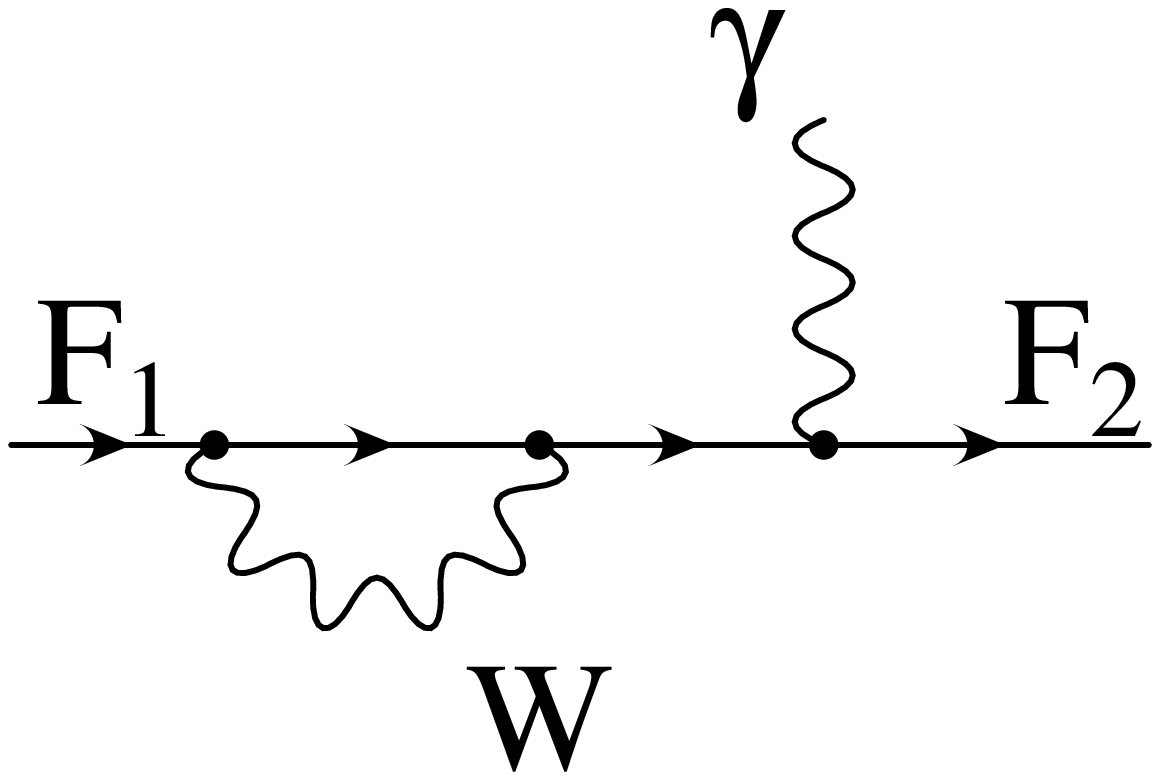,width=21mm,bbllx=210pt,%
bblly=410pt,bburx=630pt,bbury=550pt}
&\hspace*{-11mm}$+$\hspace*{4mm}&
\psfig{figure=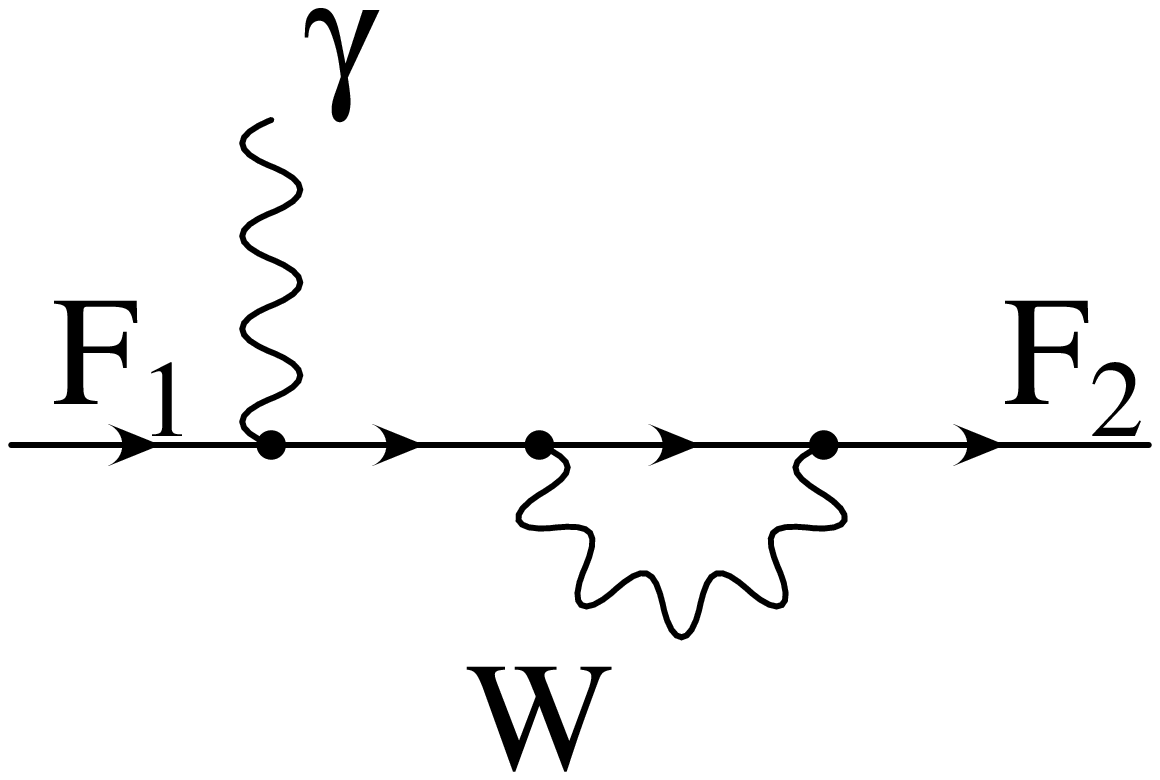,width=21mm,bbllx=210pt,%
bblly=410pt,bburx=630pt,bbury=550pt}
\\[5mm]
\end{tabular}}
\]
\end{minipage}
\begin{center}
Fig. (3)
\end{center}
\vspace*{.2cm}
%%%%%%%%%%%%%%%%%%%%%%%%%%%%%%%%%%%%%%%%%%%%%%%%%%%%%%%%%%%%

An explicit calculation \cite{S80b} of $\Gamma^{F_1 F_2\gamma}$ shows
that it is proportional to $\Delta$, the external photon momentum.  
Therefore, in the calculation of $d_e$ we can neglect
$\Delta$ in the remainder of the diagram in fig.~2a.
Also, we have $\Gamma^{F_1 F_2\gamma}=\Gamma^{F_2 F_1\gamma*}$.
These two observations are enough to show that the diagram (2a) is
complex conjugate of the diagram (2b).  The imaginary part of their
sum vanishes and no EDM is generated.  This finishes the original
proof \cite{S80}.

We would like to remark that interesting arguments in favor of 
vanishing of the two-loop contributions to EDM were also made in
\cite{Kr86,Chang}.  

\section{Explicit Two-Loop Calculation}
Our proof which we present in this section can be summarized
by saying: the diagram in fig.~(2a) vanishes.  In other words, for the
vanishing of EDM we need not examine the sum (2a)+(2b) because
they are zero independently.

The contribution  of the diagram (2a) to EDM  
is obtained from the sum of four
diagrams.  Its vanishing can be seen from the two equalities displayed
in fig.~4, which we established by an explicit calculation of the
two-loop diagrams (the insertions of the external photon are indicated
by $\otimes$.)

%%%%%%%%%%%%%%%%%%%%%%%%%%%%%%%%%%%%%%%%%%%%%%%%%%%%%%%%%%%%
\vspace{5mm}
\begin{minipage}{12.cm}
\hspace*{.8cm}
\[
\mbox{
\hspace*{-5mm}
\begin{tabular}{cccccc}
\hspace*{2mm}
\psfig{figure=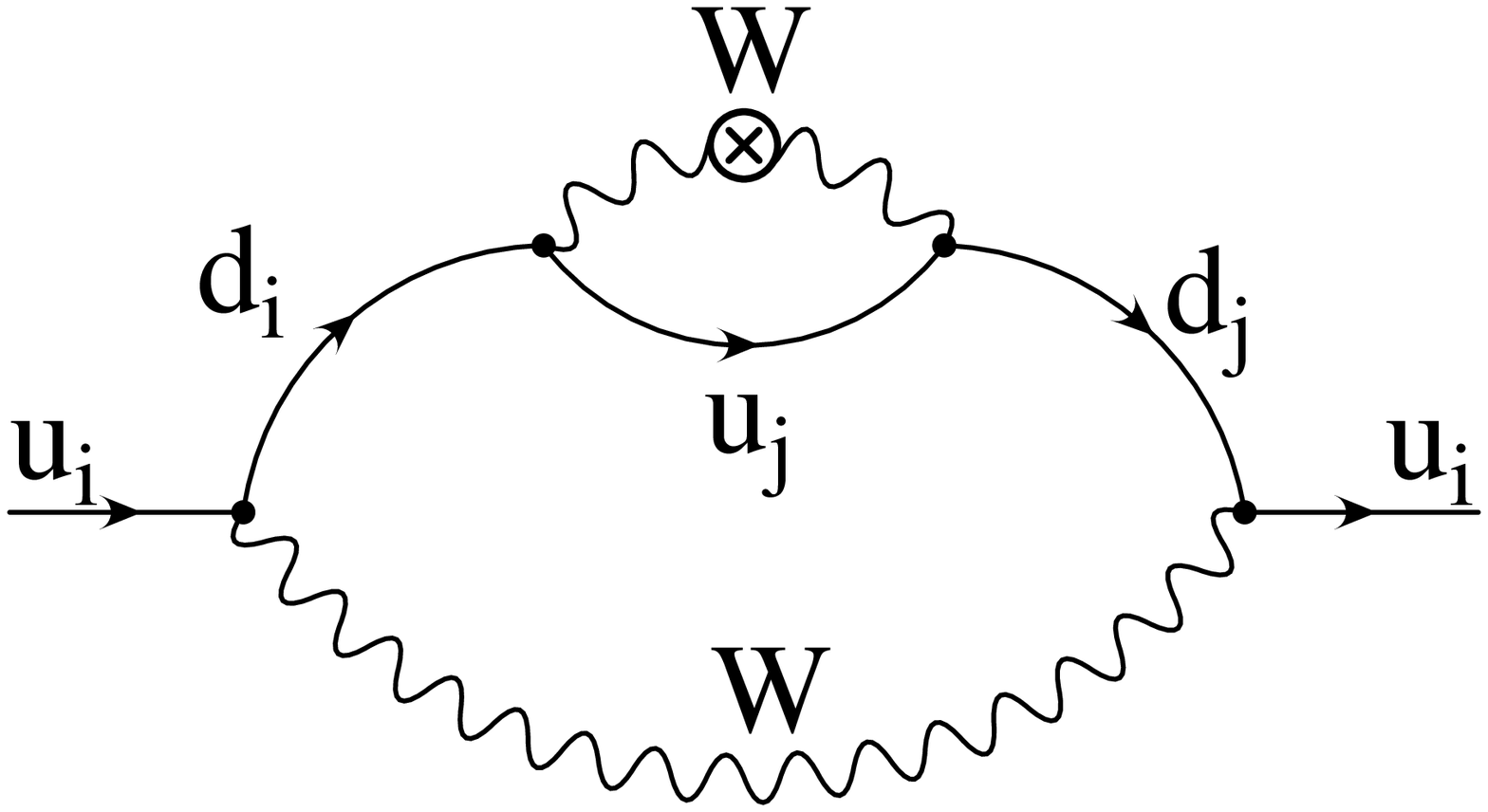,width=25mm,bbllx=210pt,%
bblly=410pt,bburx=630pt,bbury=550pt} 
\hspace*{-6mm}
&$\hspace*{-2mm}=\:-{3}\Bigg[ $ &
\hspace*{4mm}
\psfig{figure=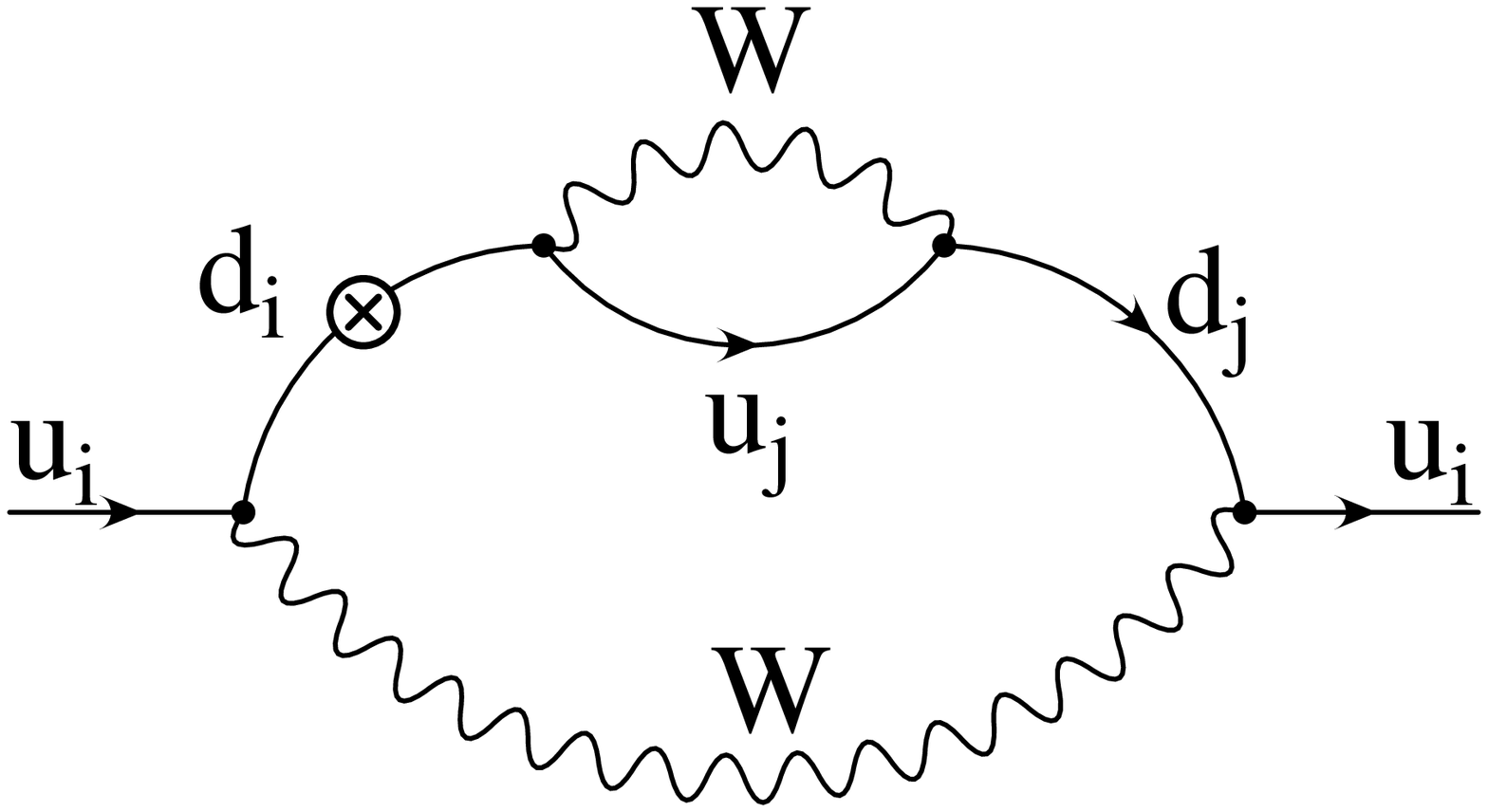,width=25mm,bbllx=210pt,%
bblly=410pt,bburx=630pt,bbury=550pt}
&\hspace*{-6mm}+\hspace*{8mm}&
\psfig{figure=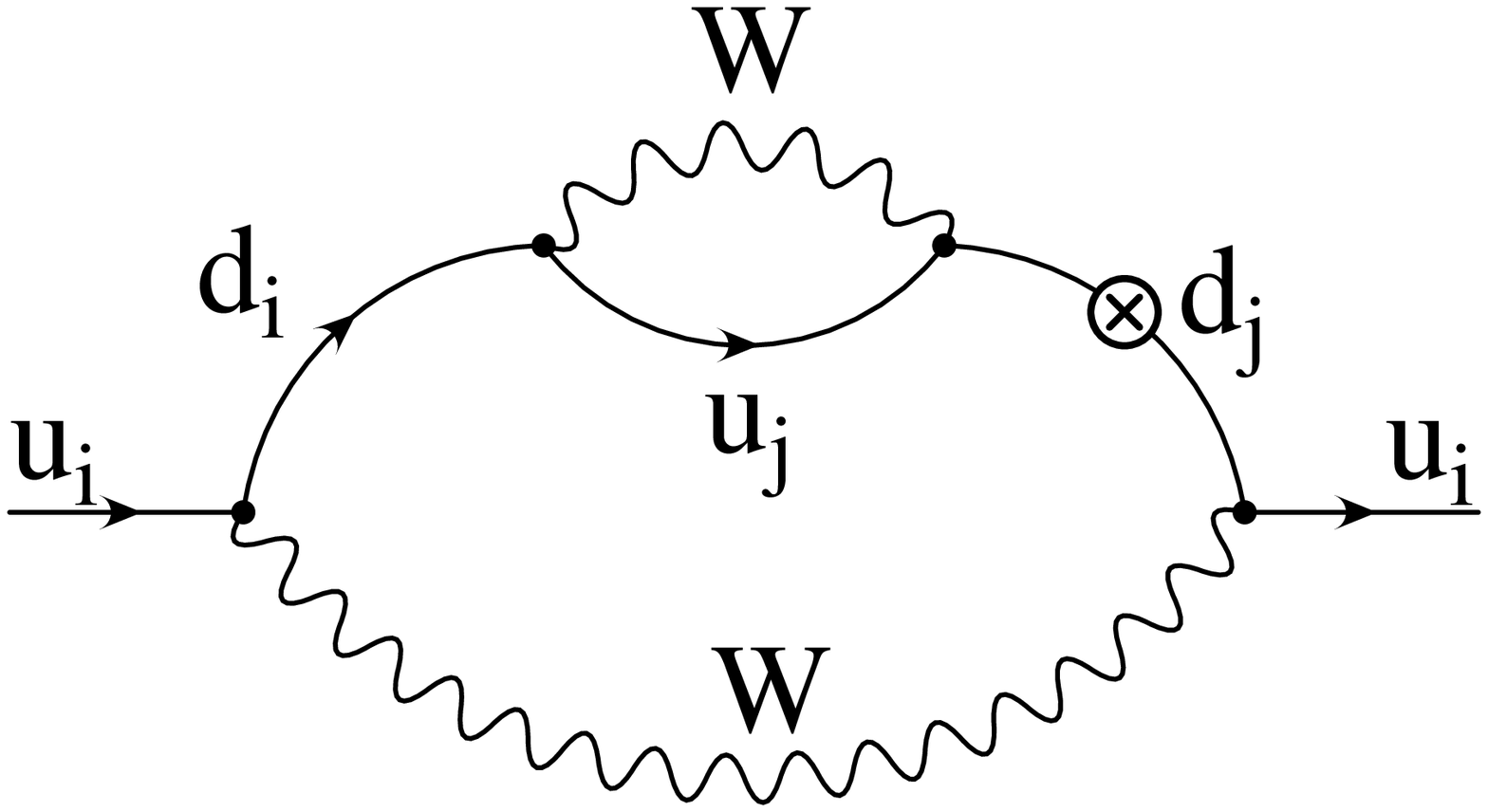,width=25mm,bbllx=210pt,%
bblly=410pt,bburx=630pt,bbury=550pt}
&\hspace*{-5mm}$\hspace*{-5mm} \Bigg]$\\[10mm]
&
$\hspace*{2mm}=\hspace*{2mm}-{3\over 2}$\hspace*{3mm}&
\hspace*{2mm}
\psfig{figure=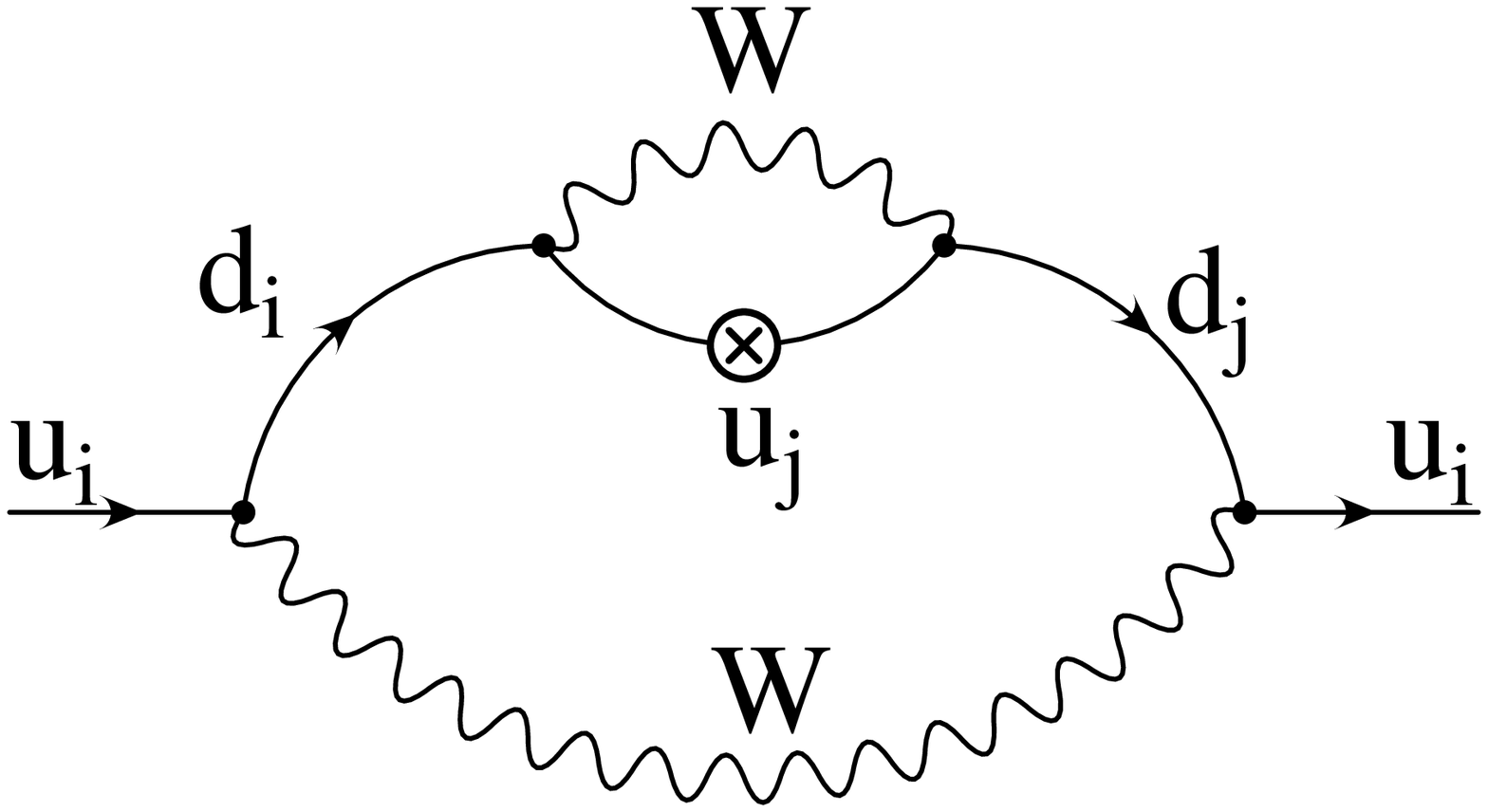,width=25mm,bbllx=210pt,%
bblly=410pt,bburx=630pt,bbury=550pt}&&&
\\[10mm]
\end{tabular}}
\]
\end{minipage}
\begin{center}
Fig. (4)
\end{center}
\vspace*{.2cm}
%%%%%%%%%%%%%%%%%%%%%%%%%%%%%%%%%%%%%%%%%%%%%%%%%%%%%%%%%%%%

The above equalities are true only in the unitary gauge.  In a
different gauge, e.g.~the linear 't Hooft--Feynman gauge, there are
additional diagrams with Goldstone bosons; in that situation only a
sum over all possible photon insertions {\em and} all flavors of the
fermion in the internal loop vanishes.  (Still, the vanishing occurs
for a fixed order of the quarks $d_i$ and $d_j$ --- no summation over
the mirror reflections is necessary.)

There is a simple relation between the contributions to the EDM form
factor of a diagram and its mirror reflection: 
for a given photon insertion the reflection
gives a minus sign and complex conjugation of the CKM factors (this is
different from the contributions to the {\em magnetic} moment --- there
is no sign change there.)    
As a result the
sum of a pair of mirror diagrams has {\em twice} the imaginary part of
an individual diagram (only the real part vanishes).

Finally, we notice that the original argument summarized in 
section 2 did not say anything about the vanishing of the real
part of the sum of the mirror diagrams; a non-vanishing result would
contribute to the imaginary part of $d_e$, forbidden by the
hermiticity of the Hamiltonian.   From the equalities we have
displayed above it is clear that both real and imaginary parts
vanish. 

%%%%%%%%%%%%%%%%%%%%%%%%%%%%%%%%%%%%%%%%%%%%%%%%%%%%%%%%%%%%

\section{Projection Operator for EDM Form \protect\\ {Factor}}

In calculations of the anomalous magnetic moment the standard
technique is to
project out the relevant Lorentz structure and average over the directions
of the external photon momentum.  The problem is then reduced to a
calculation of scalar
propagator integrals.  Such calculations are most easily performed
using algebraic manipulation programs.  

In this section we derive a projector also
for the electric dipole moment operator; the derivation follows the
case of the magnetic moment \cite{Rem,CK96}. 
We consider the most general matrix element of a
current between spin 1/2 fermions
\be
\langle\alpha_f|M_\mu|\alpha_i\rangle &=& 
{\bar u}_f(p_2)\left[F_1(t) \gamma_\mu
-{i\over 2m}F_2(t)\sigma_{\mu\nu}\Delta^\nu
+{1\over m}F_3(t) \Delta_\mu  \right. \nonumber\\
& &\hspace{-3cm}\left.+ \gamma_5\left(G_1(t) \gamma_\mu
-{i\over 2m}G_2(t)\sigma_{\mu\nu}\Delta^\nu
+{1\over m}G_3(t) \Delta_\mu \right)\right] u_i(p_1) 
\ee
with $\Delta = p_1-p_2$ and $t=\Delta^2$. 
For on-shell external fermions we have 
\be
p_1^2 = p_2^2 = m^2
\ee
We introduce $p = {1\over2}(p_1+p_2)$ for which we find
\be
p^2 = {1\over4}(4 m^2-t)\:, \hspace{1cm} p\cdot\Delta = 0 \, .
\ee

Conservation of the electromagnetic current requires $F_3(t) = 0$. 
$F_1(t)$ is the charge
form factor, $F_2(t)$ the magnetic moment form factor, $G_2(t)$ the 
electric moment form factor.
The electric dipole moment $d_e$ of the fermion is given by
\be
d_e  = G_2(0)
\ee
In order to extract the electric dipole moment form factor one can
introduce a projection operator
\be
L_\mu = (\psla_1+m)\gamma_5\left[g_1\gamma_\mu-{1\over m}g_2 p_\mu
        -{1\over m}g_3 \Delta_\mu \right](\psla_2+m) \: .
\ee

In order to determine the coefficients of $g_i$ we take the trace of
$L_\mu M^\mu$ (we work in $d$ dimensions):
\be
{\rm Tr}(L_\mu M^\mu) &=& 
    \left\{-\left[8 m^2(d-1)+2t(2-d)\right]g_1 - 4tg_3\right\}G_1(t) 
\nonumber\\
&& +\left[g_2 t\left(2-{t\over2 m^2}\right)\right]G_2(t) \nonumber\\
&& +\left[-4g_1 t -2 {t^2\over m^2}g_3\right] G_3(t)
\ee
The resulting system of equations for $G_2(t)$ can be solved by 
choosing $g_1=g_3=0$ and
\be
g_2 = {2m^2\over t(4m^2-t)}\; .
\ee
We can write the EDM form factor $G_2(t)$ as
\be
G_2(t) = {2mp^\mu \over t (t-4m^2)}{\rm Tr}\left[
  (\psla_1+m)\gamma_5
         (\psla_2+m) M_\mu \right]
\ee
Since we are only interested in the special case $t=0$ we can
make further simplifications.  As a
first step the general amplitude $M_\mu$ can then be expanded to first
order in $\Delta_\mu$:
\be
M_\mu(p,\Delta) \approx M_\mu(p,0)
+\left. \Delta_\nu{\partial\over\partial\Delta_\nu}
M_\mu(p,\Delta)\right|_{\Delta=0}
\!\!\! \equiv V_\mu(p)+\Delta^\nu T_{\nu\mu}(p) \, .
\ee
The next step is to average over the spatial directions of $\Delta$
with the formulas
\be
\langle\Delta_\mu\Delta_\nu\rangle &=& 
{1\over d-1}\Delta^2\left(g_{\mu\nu}-{p_\mu p_\nu\over p^2}\right)\,
,\nonumber\\
\langle\Delta_\mu\rangle &=& 0 \; ; 
\ee
after that, the limit $t\to 0$ can be taken. The result is
\be
\!\!\!\!\!d_e\!\!\! &=&\!\!\! 
-{p_\mu\over 4m} {\rm Tr}\left\{ \gamma_5  V^\mu
+{1\over 2m^2(d-1)}(\psla+m)\gamma_5[\psla,\gamma_\nu](\psla+m)T^{\nu\mu}
\right\}\,\,
\ee

\section{Acknowledgement}
This research was supported by BMBF 057KA92P and by 
``Gra\-duier\-ten\-kolleg
Elementarteilchenphysik'' at the University of Karlsruhe.

\end{document}